\documentclass[titlepage,12pt,a4paper,reqno]{article}
\usepackage{amsmath,amssymb,amsfonts,amsthm}
\usepackage{epsfig,epstopdf,titling,url,array,graphics}
\usepackage{bbold}
\usepackage[linktocpage=true,colorlinks,pdftex]{hyperref}
\usepackage{xcolor}
\colorlet{linkequation}{blue}
\usepackage{bm}
\usepackage{doi}
\usepackage{hyperref}
\hypersetup{colorlinks, citecolor=violet, filecolor=black, linkcolor=black, urlcolor=blue}
\newcommand*{\refeq}[1]{%
  \begingroup
    \hypersetup{
      linkcolor=linkequation,
      linkbordercolor=linkequation,
    }%
    \ref{#1}%
  \endgroup
}
\allowdisplaybreaks

\theoremstyle{plain}

\theoremstyle{definition}
\newtheorem{defn}{Definition}[section]

\theoremstyle{remark}

\begin{document} 


\begin{titlepage}

\centerline{\LARGE \bf Field Line Solutions}
\medskip
\centerline{\LARGE \bf of the Einstein-Maxwell Equations}
\vskip 1cm
\centerline{ \bf Ion V. Vancea }
\vskip 0.5cm
\centerline{\sl Grupo de F{\'{\i}}sica Te\'{o}rica e Matem\'{a}tica F\'{\i}sica}
\centerline{\sl Departamento de F\'{\i}sica}
\centerline{\sl Universidade Federal Rural do Rio de Janeiro}
\centerline{\sl Serop\'{e}dica - Rio de Janeiro, Brazil}
\centerline{
\texttt{\small ionvancea@ufrrj.br} 
}

\vspace{0.5cm}

\centerline{11 April 2020}

\vskip 1.5cm

\centerline{\large\bf Abstract}

In this paper, we are going to review the gravitating electromagnetic field in the 1+3 formalism on a general hyperbolic space-time manifold. We also discuss the recent results on the existence of the local field line solutions of the Einstein-Maxwell equations that generalize the Ra\~{n}ada solutions from the flat space-time. The global field line solutions do not always exist since the space-time manifold could impose obstructions to the global extension of various geometric objects necessary to build the fields. One example of a gravitating field line solution is the Kopi\'{n}ski-Nat\'{a}rio field which is discussed in some detail. 

\vskip 0.4cm 

\noindent \textbf{Keywords:} {Einstein-Maxwell equations. Local field line solutions. 
Generalization of Ra\~{n}ada solutions.}

\noindent

\noindent 

\end{titlepage}


\section{Introduction}

Recently, there has been an increasing interest in the topological aspects of the electromagnetism. This research line was pioneered by Trautman and Ra\~{n}ada in the seminal papers \cite{Trautman:1977im,Ranada:1989wc,Ranada:1990}. Since then, many developments and applications of the topological electromagnetic fields have appeared in the literature, ranging from the atomic particles physics to the colloidal matter physics. The convergence of classical electrodynamics and topology is mutually beneficial for both fields. For a recent list of references and an updated review of the topological solutions in the classical electrodynamics, we refer to \cite{Arrayas:2017sfq,Vancea:2019zdl}.

Most of the research on the topological electromagnetic fields has been done in the context of the classical electrodynamics applied to low energy phenomena or in the vacuum. However, the electromagnetic fields have a larger range of applications which include cosmological and astrophysical processes that usually take place at high energy. Generalizing the topological electromagnetism to these systems poses new challenges from the physical and mathematical point of view since in these cases one has to deal with the gravitating electromagnetic fields. Very recently, important steps in analyzing the topological properties of gravitons as well as of higher spin particles have been taken in \cite{Dalhuisen:2012zz,Swearngin:2013sks,Thompson:2014pta,Thompson:2014owa,Alves:2018wku}. These works are important for the topological electromagnetism since not only they generalize the topological properties of the classical photons but also contain new information on spin-2 field viewed as a particular case of the higher spin fields. However, the analyses from \cite{Dalhuisen:2012zz,Swearngin:2013sks,Thompson:2014pta,Thompson:2014owa,Alves:2018wku} focus on spin-2 fields in flat space-time. Therefore, they do not address the problem of gravitating topological electromagnetism. That problem is studied in \cite{Vancea:2017tmx}, where it was proved that there are local field line solutions of Einstein-Maxwell equations on hyperbolic space-time manifolds that are a direct generalization of Ra\~{n}ada's solutions. Another contribution to the gravitating topological electromagnetism can be found in \cite{Kopinski:2017nvp}, where a particular topological solution of the Einstein-Maxwell equations in the Einstein universe was given. Also, it is worthwhile mentioning the work \cite{Silva:2018ule}, in which the electromagnetic knots on hyperbolic manifolds are treated formally from the point of view of the foliation theory. 
In \cite{Lechtenfeld:2017tif,Kumar:2020xjr}, a new important 
method of finding the electromagnetic knot solutions in Minkowski space, by completely solving the Maxwell equations first in de Sitter space and then conformally mapping to Minkowski space was presented.

In the present chapter, we are going to give a pedagogical introduction to the problem of the topological gravitating electromagnetic fields. Since the space-time assumes the properties of a curved manifold in the presence of gravity, there are difficulties in analysing 
the gravitating fields inherited from the general formulation of the field theories on curved manifolds. While in the flat space-time, the topological electromagnetic solutions of the linear Maxwell equations can be calculated in the vacuum, in the presence of the gravitational field, the vacuum contains components of the metric that do not decouple from the electromagnetic field. In this case, the topological fields like knots and tori are better described in terms of electric and magnetic field lines which make more transparent their geometrical and topological properties. However, the decomposition of the electromagnetic field in electric and magnetic components is not possible on general curved manifolds. One remarkable exception are the hyperbolic manifolds for which there is a canonical formulation of the General Relativity, called \emph{1+3 - formalism}, that allows one to put the Einstein-Maxwell equation in a form similar to the Maxwell equations in the Minkowski space-time \cite{Gourgoulhon:2012}. By using the $1+3$ - formalism, it has been shown in \cite{Vancea:2017tmx} that Ra\~{n}ada's solutions have a \emph{local generalization} to hyperbolic manifolds. However, one cannot expect \emph{global solutions} to exist on general hyperbolic manifolds. Indeed, global field line solutions depend on the global properties of the space-time manifold that could contain obstructions to the global extension of certain mathematical objects such as the differential forms. Therefore, the global solutions are remarkable and they are associated to particular manifolds. This is the case of the static Einstein universe for which a topological radiation field that depends only on time was found in \cite{Kopinski:2017nvp}. 

This work is organized as follows. In Section 2, we are going to review the basic features of the Maxwell equations in flat space-time in terms of differential forms. Since our main focus is the generalization of Ra\~{n}ada's solutions to the gravitating electromagnetic field, we will introduce in Section 3 the $1+3$ - formalism which makes the correspondence between the equations of motion of the electromagnetic field in flat and in curved space-time, respectively, more transparent. This formalism will be used to write the Einstein-Maxwell equations on hyperbolic manifolds. In Section 4, we will revisit the proof of existence of local field line solutions of Einstein-Maxwell equations on hyperbolic manifolds from \cite{Vancea:2017tmx}. In Section 5, the Kopi\'{n}ski-Nat\'{a}rio solution in the Einstein space is presented. We collect some useful mathematical definitions in the Appendix. The units used throughout this paper are natural with $c=G=1$. 

\section{Field line solutions in flat space-time}

In this section, we are going to survey the Maxwell equations in the Minkowski space-time and establish our notations. The material presented here is well-known and can be found in standard texts on classical electrodynamics, see e. g. \cite{Jackson:1998nia}. We also review the Ra\~{n}ada field line solutions from \cite{Ranada:1989wc,Ranada:1990}. A more detailed review can be found in a different chapter from this volume \cite{Vancea:2019zdl}.

\subsection{The Maxwell's equations}

Consider the Minkowski space-time $\mathbb{R}^{1,3} = \{ \mathbb{R}^{4}, \mathrm{\eta} \}$, where $\mathrm{\eta}$ is the Minkowski pseudo-metric tensor of signature $(-, +, + , +)$. The events from $\mathbb{R}^{1,3}$ are identified by the coordinate four-vectors $x^{\mu} = (t, \mathbf{x} ) = (t, x^i)$, where the indices $i, j, k = 1, 2, 3$ denote the space-like components. The electromagnetic field on $\mathbb{R}^{1,3}$ can be described in terms of the four-vector potential $A^{\mu} = (\phi , \mathbf{A})$, where $\phi$ is the scalar potential and $\mathbf{A} = (A^1 , A^2, A^3)$ is the three-dimensional vector potential. Then the electromagnetic field is given by the rank-2 antisymmetric tensor $F_{\mu \nu}$ defined by the following relation
\begin{equation}
F_{\mu \nu} = \partial_{\mu} A_{\nu} - \partial_{\nu} A_{\mu} \, .
\label{EM-field}
\end{equation}
The tensor $F_{\mu \nu}$ has six independent degrees of freedom which are identified with the components of the electric and magnetic vector fields as follows
\begin{equation}
E_{i} = \partial_i A_0 - \partial_0 A_i \, ,
\qquad
B_{i} = \varepsilon_{ijk} \partial_j A_k \, .
\label{E-M-components}
\end{equation}
The Hodge dual (pseudo)-tensor associated to $F_{\mu \nu}$ is defined by the following relation
\begin{equation}
{\star F}_{\mu \nu} = \frac{1}{2} \varepsilon_{\mu \nu \rho \sigma} F^{\rho \sigma} \, ,
\qquad
F^{\mu \nu} = \eta^{\mu \rho} \eta^{\nu \sigma} F_{\rho \sigma} \, .
\label{F-Hodge-dual}
\end{equation}
One can write the electromagnetic field in terms of differential forms on  $\mathbb{R}^{1,3}$. This formulation will be useful later when we will discuss the Einstein-Maxwell equations. Let us introduce the 2-form electromagnetic field $F$ in the three-dimensional and four-dimensional representations, respectively, given by the following relations 
\begin{align}
F & = B + E \wedge d x^0 \, ,
\label{F-two-form-1}
\\
F & = F_{\mu \nu} \, dx^{\mu} \wedge dx^{\nu} \, ,
\label{F-two-form-2}
\end{align}
where $B$ is the magnetic 2-form, $E$ is the electric 1-form and $\wedge$ is the wedge product of differential forms. Then the Hodge star operator is defined by the following relations
\begin{align}
\star \,  : \, \Omega^{k} (\mathbb{R}^{1,3}) & \to \Omega^{4-k} (\mathbb{R}^{1,3}) \, ,
\label{Hodge-dual-1}
\\
\omega \wedge \left( \star \sigma \right) & = \langle \omega , \sigma \rangle \mathrm{n} 
\, ,
\label{Hodge-dual-2}
\end{align}
where $\mathrm{n}$ is an unitary vector and $\langle \cdot , \cdot \rangle$ is the scalar product. Here, we denote by $\Omega^{k} (M)$ the linear space of $k$-forms on the manifold $M$. The Maxwell equations in terms of differential forms are given by just two differential equations
\begin{align}
d F & = 0 \, ,
\label{Maxwell-form-1}
\\
\star \, d \star F & = J \, , 
\label{Maxwell-form-2}
\end{align}
where $J = J_{\mu} dx^{\mu}$ is the 1-form current density \cite{Frankel:1997ec}. The equations (\refeq{Maxwell-form-1}) and (\refeq{Maxwell-form-2}) are covariant which means that their  symmetry group is the Poincar\'{e} group.

In what follows, we are going to discuss a particular type of solutions of Maxwell's equations in the vacuum which is defined by the absence of currents, namely, 
\begin{align}
d F & = 0 \, ,
\label{Maxwell-form-1-vacuum}
\\
\star \, d \star F & = 0\, , 
\label{Maxwell-form-2-vacuum}
\end{align}
In this case, the Maxwell equations have an additional symmetry which is the electromagnetic duality given by the interchange of the forms $F$ and $\star F$ with each other
\begin{equation}
F \leftrightarrow \star F \, .
\label{EM-duality}
\end{equation}
It is important to recall a property of the Hodge star operator. If $\omega$ is an arbitrary $k$-dimensional form on a manifold $M$ of dimension $n$, then
\begin{equation}
\star \star \, \omega = (-1)^{k(n-k)} \mbox{sign}\left(\det{\eta} \right) \omega \, .
\label{Hodge-star-property-n-k}
\end{equation}
It follows from the equation (\ref{Hodge-star-property-n-k}) that if $M = \mathbb{R}^{1,3}$ and $k=2$, the Hodge star operator satisfies the equation
\begin{equation}
\star^2  = - 1 \, .
\label{Hodge-star-property-4-2}
\end{equation}
From the equation (\refeq{Hodge-star-property-4-2}), one can deduce that the operator $\star^2$ has the eigenvalues $\pm i$ when acting on the 2-forms. Therefore, the fields $F$ belong to one of the two classes of self-dual or anti-self-dual 2-forms. Due to the linearity of Maxwell's equations (\refeq{Maxwell-form-1-vacuum}) and (\refeq{Maxwell-form-2-vacuum}), a general solution in vacuum is a superposition of solutions from each class
\begin{equation}
F = F_{+} + F_{-} \, ,
\qquad
\star F_{\pm} = \pm i F \, .
\label{F-form-pm}
\end{equation}
The above equations shows that $F$ is a complex 2-form. It is a simple exercise in electrodynamics to prove that the equations (\refeq{Maxwell-form-1-vacuum}) and (\refeq{Maxwell-form-2-vacuum}) reproduce the three-dimensional Maxwell equations. This can be seen by decomposing the differential forms defined on the Minkowski space-time with respect to its global spatial foliation whose leaves are isomorphic to $\mathbb{R}^3$. The derived  electromagnetic forms $d F$ and $\star \, d \star F$ have the following decomposition
\begin{align}
d F & = d B + d E \wedge d x^0 \, ,
\label{Maxwell-4d-1-a}
\\
\star \, d \star F & = - \partial_0 E 
- \boldsymbol{\star} \, \mathbf{d} \boldsymbol{\star} E \wedge d x^0 
+ \boldsymbol{\star} \, \mathbf{d} \boldsymbol{\star} B \, ,
\label{F-dual-decompose}
\end{align}
where $\mathbf{d}$ is the three-dimensional differential derivative and $\boldsymbol{\star}$ is the three-dimensional Hodge star operator. Then the Maxwell equations take the following form
\begin{align}
\mathbf{d} E + \partial_0 B & = 0 \, ,
\label{Maxwell-4d-1-E}
\\
\mathbf{d} B & = 0 \, ,
\label{Maxwell-4d-1-B}
\\
\boldsymbol{\star} \, \mathbf{d} \boldsymbol{\star} E & = \rho \, ,
\label{Maxwell-4d-2-E}
\\
\boldsymbol{\star} \, \mathbf{d} \boldsymbol{\star} B 
- \partial_0 E 
& = \mathbf{J} \, ,
\label{Maxwell-4d-2-B}
\end{align}
where $\mathbf{J} = J_i dx^i$. In the vacuum, the equations (\refeq{Maxwell-4d-2-B}) have a  simpler form
\begin{align}
\mathbf{d} E + \partial_0 B & = 0 \, ,
\label{Maxwell-4d-3-E}
\\
\mathbf{d} B & = 0 \, ,
\label{Maxwell-4d-3-B}
\\
\mathbf{d} \boldsymbol{\star} E & = 0 \, ,
\label{Maxwell-4d-4-E}
\\
\boldsymbol{\star} \, \mathbf{d} \boldsymbol{\star} B 
- \partial_0 E 
& = 0 \, .
\label{Maxwell-4d-4-B}
\end{align}
The formalism of differential forms allows one to calculate the properties of the electromagnetic field in a coordinate independent manner. Also, the equations obtained in this formalism are more compact. For more details we refer the reader to the chapter \cite{Vancea:2019zdl} from this volume or to the excellent book \cite{Frankel:1997ec}.

\subsection{Field line solutions}

In order to describe the dynamics of the electromagnetic field in the vacuum, the equations (\refeq{Maxwell-4d-3-E}) - (\refeq{Maxwell-4d-4-B}) must be solved with proper boundary conditions. As it is well known, the most general solution to the above set of equations 
is given by a superposition of monocromatic waves that satisfy the dispersion relation $\mathbf{k}^2 = \omega^2(\mathbf{k})$ where $\mathbf{k}$ is the wave vector and $\omega(\mathbf{k})$ is the correspondig wave frequency in the infinite empty space $\mathbb{R}^3$ \cite{Jackson:1998nia}. Since these solutions are known for a long time, it came as a surprise when Trautman and Ra\~{n}ada published three articles in which they presented independetly of each other, a new type of solution of Maxwell's equations in vacuum with a non-trivial topological structure \cite{Trautman:1977im,Ranada:1989wc,Ranada:1990}. All solutions of the topological class are characterized by new non-zero \emph{topological charges} which are the \emph{link numbers} between the electric and magnetic field lines. Also, the topological solutions can be expressed in terms of field lines which justifies their name of \emph{field line solutions}. A more extended review of the field line solutions is presented in the chapter \cite{Vancea:2019zdl} from this volume. Here, we are only going to recall the general form of Ra\~{n}ada's solutions which is relevant for the rest of our discussion.

The Ra\~{n}ada solutions are described in terms of two smooth complex scalar fields $\phi$ and $\theta$ in the Minkwoski space-time \cite{Arrayas:2017sfq}
\begin{equation}
\phi : \mathbb{R}^{1,3} \rightarrow \mathbb{C} \, , 
\qquad
\theta : \mathbb{R}^{1,3} \rightarrow \mathbb{C} \, . 
\label{Ranada-complex-fields-1}
\end{equation}
The main role of these fields is to serve as a backbone for the topology of the field lines in $\mathbb{R}^3$ in the following sense: the electric and magnetic field lines are the level curves of $\phi$ and $\theta$, respectively. Then one can show that all solutions of Maxwell's equations that have this property are of the following form
\begin{align}
F_{\mu \nu} & = g(\bar{\phi} , \phi ) 
\left( 
\partial_{\mu} \bar{\phi} \, \partial_{\nu} \phi
-
\partial_{\nu} \bar{\phi} \, \partial_{\mu} \phi
\right) \, ,
\label{Ranada-magnetic-solution-1}
\\
\star F_{\mu \nu} & = f(\bar{\theta} , \theta ) 
\left( 
\partial_{\mu} \bar{\theta} \, \partial_{\nu} \theta
-
\partial_{\nu} \bar{\theta} \, \partial_{\mu} \theta
\right) \, .
\label{Ranada-electric-solution-1}
\end{align}
The field line solutions given by the equations (\refeq{Ranada-magnetic-solution-1}) and (\refeq{Ranada-electric-solution-1}) are parametrized by two smooth function $g$ and $f$ that depend on $\theta$ and $\phi$. The electromagnetic differential forms are related to the tensors $F_{\mu \nu}$ and $\star F_{\mu \nu}$ by the following relations
\begin{align}
F & = 
-\varepsilon_{j k l} B_{j} dx^{k} \wedge dx^{l} 
+ E_{j} dx^{j} \wedge d x^{0} \, , 
\label{F-field-components-1}
\\
\star F & = 
\varepsilon_{j k l} E_{j} dx^{k} \wedge dx^{l} 
+ B_{j} dx^{j} \wedge d x^{0} \, . 
\label{star-F-field-components-1}
\end{align} 
A Ra\~{n}ada solution of Maxwell's equations is an electromagnetic field of the form (\refeq{Ranada-magnetic-solution-1}) and (\refeq{Ranada-electric-solution-1}) that has the following components 
\begin{align}
E_{j} & = 
\frac{\sqrt{a} }{2 \pi i} \left( 1 + |\theta|^2 \right)^{-2} 
\varepsilon_{jkl} \, \partial_{k} \bar{\theta} \, \partial_{l} \theta \, ,
\label{Ranada-E-1}
\\
B_{j} & = 
\frac{\sqrt{a}}{2 \pi i} \left( 1 + |\phi|^2 \right)^{-2} 
\varepsilon_{jkl} \, \partial_{k} \bar{\phi} \, \partial_{l} \phi \, .
\label{Ranada-B-1}
\end{align}
Here, $a$ is a parameter that fixes the physical dimension of the solution. The fields $\mathbf{E}$ and $\mathbf{B}$ defined by the equations (\refeq{Ranada-E-1}) and (\refeq{Ranada-B-1}) correspond to particular choices of $f$ and $g$ from the equations (\refeq{Ranada-magnetic-solution-1}) and (\refeq{Ranada-electric-solution-1}). Also, we obtain the following constraints on the paramenters from the self-duality condition in three-dimensions
\cite{Arrayas:2017sfq}
\begin{align}
\left( 1 + |\phi|^2 \right)^{-2} \varepsilon_{jmn} 
\partial_m \phi \partial_n \bar{\phi}
& = \left( 1 + |\theta|^2 \right)^{-2} 
\left(
\partial_0 \bar{\theta} \partial_j \theta
-
\partial_0 \theta \partial_j \bar{\theta}
\right)
\, ,
\label{fi-teta-1-1}
\\
\left( 1 + |\theta|^2 \right)^{-2} \varepsilon_{jmn} 
\partial_m \bar{\theta} \partial_n \theta
& = \left( 1 + |\phi|^2 \right)^{-2}
\left(
\partial_0 \bar{\phi} \partial_j \phi
-
\partial_0 \phi \partial_j \bar{\phi}
\right)
\, .
\label{fi-teta-2-1}
\end{align}
Some comments are in order here. Firstly, note that the Ra\~{n}ada solutions satisfy the orthogonality property
\begin{equation}
E_j B_k \delta_{j k} = 0 \, .
\label{null-field-1-1}
\end{equation}
Secondly, since the fields given by the relations (\refeq{Ranada-magnetic-solution-1}) and (\refeq{Ranada-electric-solution-1}) satisfy the Maxwell equations, the charges associated to the Poincar\'{e} group and the $U(1)$ group are conserved. However, there are new charges associated to the topology of the field lines. The topological observables related to the topological charges are the pure electric and magnetic helicities of the electromagnetic field that are defined as follows
\begin{align}
H_{ee} & = \int d^3 x \, \delta_{ij} E_i C_j = 
\int d^3 x \, \varepsilon_{jkl} C_j \partial_k C_l \, , 
\label{helicity-Hee-1}
\\
H_{mm} & = \int d^3 x \, \delta_{ij} B_i A_j = 
\int d^3 x \, \varepsilon_{jkl} A_j \partial_k A_l \, , 
\label{helicity-Hmm-1}
\end{align}
where $\mathbf{A}$ and $\mathbf{C}$ are the corresponding magnetic and electric potential vectors 
\begin{equation}
E_j = \varepsilon_{jkl} \partial_k C_l \, ,
\qquad
B_j = \varepsilon_{jkl} \partial_k A_l \, .
\label{C-A-potentials-1}
\end{equation}
Observed that the correspondence between the helicities and the topological objects is a consequence of the definition of the field line solution in terms of the complex scalar fields 
$\phi$ and $\theta$. 

If the electromagnetic solutions carry a finite energy, then they should take zero value in the limit $|\mathbf{x}| \to \infty$. In order for that to happen, the fields $\phi (t, \mathbf{x} )$ and $\theta(t,\mathbf{x})$ must have the same asymptotic behaviour as the energy. That implies that $\phi$ and $\theta$ are complex functions on $\mathbb{R} \times S^3$ or, equivalently, they correspond to one-parameter families of maps $S^3 \to \mathbb{C}$. On the other hand, there is a natural identification $\mathbb{C} \simeq \mathbb{R}^2$. This identification can be further refined if the inverse maps $\phi^{-1}$ and $\theta^{-1}$ do not depend on the complex phases of their arguments. The functions that satisfy this property can be interpreted as maps $S^3 \to S^2$. However, these are the Hopf maps characterized by the topological number called the \emph{Hopf index}. It turns out that when this index is expressed as a Chern-Simons integral, it takes the same form as the heliticies  defined above. For a detailed discussion of these aspects, see \cite{Arrayas:2017sfq}.

\section{The $1+3$ - formalism of the General Relativity}

In this section, we review those basic concepts of the $1+3$ - formalism of the General Relativity necessary to discuss the Einstein-Maxwell equations on hyperbolic manifolds. In our presentation, we will follow mainly the reference \cite{Gourgoulhon:2012} with minor modifications of the notations.

\subsection{Space-time foliation}

According to the General Relativity, the gravity is a consequence of the non-trivial geometry of the space-time viewed as a four-dimensional differential manifold $M$ endowed with a metric tensor field $\mathbf{g} = g_{\mu \nu}$. The manifold $M$ can be approximated locally by the Minkowski space-time \cite{Weinberg:1972kfs}. It follows from the first principles that the metric $g_{\mu \nu}$ must be a solution of the Einstein equations either in the presence of the matter or with no matter at all. In this case, the dynamics of the gravitating electromagnetic field is given by a set of covariant equations on $M$. Our main goal is to discuss the existence of field line solutions of these equations. More concretely, we would like to see whether there are any local solutions of the equations of motion of the gravitating electromagnetic field that generalize the Ra\~{n}ada solutions from the flat space-time. The existence problem is not well posed globally unless further assumptions about the structure of the hyperbolic manifold are made. In the general case, there could be obstructions to the existence of certain mathematical objects, such as the 2-forms, that could prevent the existence of the global fields on $M$. 

We recall here two important properties of the mathematical structure of Maxwell's equations 
that play a silent role in the derivation of the results from the previous section:
\emph{i}) the splitting between the electric and the magnetic fields, necessary to define the electric and magnetic field lines; 
and \emph{ii}) the explicit time-evolution of the system during which the helicities are conserved. These elements can be reproduced in the presence of gravity if the manifold $M$ is globally hyperbolic, that is if it admits a foliation in terms of an one-parameter family of space-like manifolds $\Sigma \simeq \mathbb{R}^3$ parametrized by a global time $t \in \mathbb{R}$. In that case, the manifold $M$ has the topology $M = \mathbb{R} \times \Sigma$. The method of splitting the covariant equations on $M$ with respect to the foliation is called the \emph{1+3 - formalism}. In order to make the discussion more concrete, we need to introduce these concepts in a more formal way. For more details on the 1+3 - formalism see \cite{Gourgoulhon:2012}.

Let $(M,\mathbf{g})$ denote a four-dimensional space-time manifold
$M$ endowed with a smooth metric tensor field $\mathbf{g}$ of signature $(-,+,+,+)$ that has the following properties: 
\begin{itemize}
\item $M$ is time-orientable;
\item $M$ is hyperbolic.
\end{itemize}
Then there is a globally defined scalar field $\mathbf{t}$ such that $M$ is a foliation generated by $\mathbf{t}$ with the leaves defined by the following relation
\begin{equation}
\Sigma_t = \{ p \in \mathcal{M} : \mathbf{t}(p) = t = \mbox{constant} \} \, .
\label{foliation-leaves}
\end{equation}
Equivalently, one can write the foliation as $\mathcal{M} \simeq \mathbb{R}\times \Sigma$. 

Among the mathematical structures that can be defined on $M$, there are two remarkable vector fields and one scalar function that play a crucial role in the $1+3$-formalism, namely: the normal vector field $\mathbf{n}$, the normal evolution vector field $\mathbf{m}$ and the lapse function $N$. These are defined by the following relations
\begin{equation}
\mathbf{n} := -N \nabla \mathbf{t} \, ,
\qquad
\mathbf{m} := N \mathbf{n} \, ,
\qquad
N := 
\left[
- g_{\mu \nu}\nabla^{\mu} \mathbf{t} \nabla^{\nu} \mathbf{t}
\right]^{-\frac{1}{2}} \, .
\label{normal-normal-evolution-Lapse}
\end{equation}
Since we want to identify the leaves $\Sigma_t$ at every value of the parameter $t$ and at every point $p$ with the space-like submanifold of $M$, and we want the vector field $\mathbf{t}$ to be the time vector field throughout the entire $M$, the gradient $\nabla \mathbf{t}$ must be time-like. 

In order to be able to make physical measurements at a point $p \in \Sigma_t$ one needs local coordinates. One possibility is to choose $x^{\mu} = (t,x^i)$ adapted to the foliation. Here, the Latin indices $i,j = 1, 2, 3$ are for the components on the leaf. The time evolution of a physical system that contains the event $p$ at $t$ takes place along the field line of the vector field $\pmb{\partial}_{t}$ defined by the following relations  
\begin{equation}
\pmb{\partial}_{t} := \mathbf{m} + \pmb{\beta},
\, \, \, \, \,
g_{\mu \nu} \beta^{\mu} n^{\nu} = 0,
\label{relations-adapted-coordinates}
\end{equation}
where $\pmb{\partial}_{t}$ is the derivative along the adapted time and $\pmb{\beta} \in \mathcal{T}_p (\mathcal{M})$ is the shift vector corresponding to the displacement of the origin of the space-like coordinates between two infinitesimally closed leaves.

The above decomposition of the manifold $M$ is in agreement with its topological structure as a foliation. All geometrical objects defined on $M$ can be decomposed in the same way. For example, the four-dimensional metric tensor $\mathbf{g}$ on $M$ induces a three-dimensional metric $\pmb{\gamma}$ on each leaf $\Sigma_t$ by the following canonical reduction operation
\begin{equation}
\pmb{\gamma} = \mathbf{g}|_{\Sigma_t} \,
\Longleftrightarrow \,
\gamma_{ij} = g_{ij} \, .
\label{foliation-induced-metric}
\end{equation}
The equation (\refeq{foliation-induced-metric}) can be interpreted as the action of a projector ${P^{\mu}}_{\nu}$ on the tensor $\mathbf{g}$ with
\begin{equation}
{P^{\mu}}_{\nu} := {g^{\mu}}_{\nu} + n^{\mu}n_{\nu} \, ,
\qquad
P_{\mu i} \, n^{\mu} = 0 \, .
\label{projector-definition}
\end{equation}
This second interpretation is more operational as it allows one to project other mathematical objects onto $\Sigma_t$. 

If we denote the components of the covariant derivative on the leave by $D_i$, the three-dimensional connection associated to it is torsionless providing that the metric is compatible with the covariant derivative
$D^{i} \gamma_{ij}=0$.  Since the leaf $\Sigma_t$ is embedded into the four-dimensional manifold $M$, one can define its exterior curvature as follows
\begin{equation}
K_{ij}  : = {P^{\mu}}_{i} {P^{\mu}}_{j} \nabla_{\mu} n_{\nu} \, ,
\qquad
K:= \gamma^{ij} K_{ij} \, ,
\label{extrinsic-curvature}
\end{equation} 
where $\nabla_{\mu}$ are the components of the four-dimensional covariant derivative compatible with the metric $g_{\mu \nu}$. It is easy to see that the line element takes the following form under the metric decomposition from the equation (\refeq{foliation-induced-metric})
\begin{equation}
ds^2 = - (N dt)^2 + \gamma_{ij}
\left(
d x^i + \beta^i dt
\right)
\left(
d x^j + \beta^j dt
\right) \, .
\label{line-element-1+3}
\end{equation}
By acting with the projector ${P^{\mu}}_{\nu}$, one can decompose the covariant gravitational field and the electromagnetic field with respect to the foliation of the underlying manifold.

\subsection{Einstein-Maxwell equations}

The equations that govern the dynamics of the gravitating electromagnetic field can be obtained from the following action
\begin{equation}
S[g,A] =
- \int d^4 x {\sqrt {-g}}
\left(
\, F_{\mu \nu } \, F^{\mu \nu }
+ A_{\mu }\, j^{\mu }
\right) 
\, ,
\label{Einstein-Maxwell-action}
\end{equation}
where
\begin{align}
F_{\mu \nu} & = \partial_{\mu} A_{\nu} - \partial_{\nu} A_{\mu} \, ,
\label{Einstein-Maxwell-eq-1-covar}
\\
j^{\mu} & = \frac{1}{\sqrt{-g}} \partial_{\nu } \mathcal{D}^{\mu \nu} \, ,
\label{Einstein-Maxwell-eq-2-covar}
\\
\mathcal{D}^{\mu \nu} & = \sqrt{-g} F^{\mu \nu} \, .
\label{Einstein-Maxwell-eq-3-covar}
\end{align}
By applying the variational principle to the action (\refeq{Einstein-Maxwell-action}), one obtains the equations of motion of the electromagnetic field as well as of the gravitational field. These equations are the Einstein-Maxwell equations. In the $1+3$ - formalism presented above, the field strength tensor $F_{\mu \nu}(t,\mathbf{x})$ can be decomposed locally in to the electric field $E^{\mu} (t,\mathbf{x})$ and the magnetic field $B^{\mu}(t,\mathbf{x})$, respectively. The components of the electromagnetic tensor are given by the following relations
\begin{equation}
E_{\mu}(t,\mathbf{x}) = F_{\mu \nu}(t,\mathbf{x})n^{\nu} (t,\mathbf{x}) \, ,
\hspace{0.5cm}
B_{\mu}(t,\mathbf{x}) = \frac{1}{2} \varepsilon_{\mu \nu \sigma} (t,\mathbf{x})F^{\nu \sigma}(t,\mathbf{x}) \, ,
\label{electric-magnetic-decomposition}
\end{equation}
where $\varepsilon_{\mu \nu \sigma}(t,\mathbf{x})$ is the contracted four-dimensional Levi-Civita tensor. One can easily show that the electric and magnetic fields satisfy the following equations
\begin{equation}
E_{\mu} (t,\mathbf{x}) n^{\mu} (t,\mathbf{x}) = 0 \, ,
\hspace{0.5cm}
B_{\mu}(t,\mathbf{x}) n^{\mu} (t,\mathbf{x}) = 0 \, .
\label{electric-magnetic-fields-tangent}
\end{equation}
The equations (\ref{electric-magnetic-fields-tangent}) show that $E^{\mu} (t,\mathbf{x}) \, ,B^{\mu}(t,\mathbf{x}) \in \mathcal{T}_p(\Sigma_t)$, that is the electric and magnetic vectors are tangent to the leaf at $p$. It follows from (\ref{electric-magnetic-fields-tangent}) that the field strength has the following local form
\begin{equation}
F_{\mu \nu} = n_{\mu} E_{\nu} - n_{\nu} E_{\mu} + \varepsilon_{\mu \nu \rho \sigma} n^{\rho} B^{\sigma}.
\label{Electromagnetic-field-tensor}
\end{equation}
In what follows, we are going to discuss some properties of the Einstein-Maxwell equations. In order to simplify the notation, we will drop off the local space-time coordinates unless their explicit presence is strictly necessary. In this notation, the Einstein-Maxwell equations without sources are written as
\begin{align}
\mathcal{L}_{\mathbf{m}} E^{i} - NKE^{i} - 
\varepsilon^{ijk} D_{j} \left( N B_{k} \right) & = 0,
\label{Faraday-gravitation}
\\
\mathcal{L}_{\mathbf{m}} B^{i} - NKB^{i} + 
\varepsilon^{ijk} D_{j} \left( N E_{k} \right) & = 0,
\label{Ampere-gravitation}
\\
D_{i}E^{i} &= 0,
\label{Gauss-E-gravitation}
\\
D_{ i} B^{i} &= 0.
\label{Gauss-B-gravitation}
\end{align}
Note that the equations (\refeq{Faraday-gravitation}) and (\refeq{Ampere-gravitation}) 
describe the dynamics of the electromagnetic field and they correspond to the Faraday and Amp\`{e}re laws, respectively. The equations (\refeq{Gauss-E-gravitation}) and (\refeq{Gauss-B-gravitation}) are constraints on the components $E^i$ and $B^i $ and generalize the Gauss law to the gravitating electromagnetic field.

\section{Existence of local field line solutions}

The $1+3$ - formalism introduced in the previous section allows us to separate the covariant electromagnetic field in to electric and the magnetic components. This decomposition is useful if one wants to generalize the field line solutions from the Minkowski space-time to hyperbolic manifolds. In the most general case of an arbitrary hyperbolic space-time manifold $(M,\mathbf{g})$, the field line solutions are local. Nevertheless, 
global field line solutions could exist in particular case. In the rest of this section, we will review the arguments from \cite{Vancea:2017tmx} where it was given the proof of existence of local field line solutions of Einstein-Maxwell equations on general hyperbolic manifolds.

\subsection{Field line solutions}

In order to find magnetic field line solutions of the Einstein-Maxwell equations, we need to analyse only the equations (\refeq{Ampere-gravitation}) and (\refeq{Gauss-B-gravitation}). Since the problem is local, one must work within an (arbitrary) neighbourhood $U_p \in \mathcal{M}$, where $p \in \Sigma_t$. Also, it is necessary to use the adapted coordinates $(t,\mathbf{x})$ in $U_p$ as discussed before.  

As we have seen in the previous section, the magnetic field lines are defined in terms of scalar fields in the Minkowski space-time. That suggests that a scalar field $\phi : U_p \rightarrow \mathbb{C}$ be introduced on $U_p \in M$. The equation (\refeq{Ampere-gravitation}) 
is the equation of motion of $B^{i}(t,\mathbf{x})$ which should also obey the constraint (\refeq{Gauss-B-gravitation}) at all times. The form of this constraint suggests the following ansatz for the magnetic field
\begin{equation}
B^{i} (t,\mathbf{x}) = f(t,\mathbf{x}) \varepsilon^{ijk} (t,\mathbf{x}) D_{j} \phi (t,\mathbf{x}) D_{k} \bar{\phi} (t,\mathbf{x}).
\label{magnetic-field-ansatze}
\end{equation}
Here, $f(t,\mathbf{x})$ is a smooth arbitrary field on $U_p$. By plugging $B^{i} (t,\mathbf{x})$ into the equation (\refeq{Gauss-B-gravitation}), we can verify that any function $f(t,\mathbf{x})$ that depends on the space-time coordinates only implicitly via the scalar field, i. e. $f(\phi(t,\mathbf{x}),\bar{\phi}(t,\mathbf{x}))$, satisfies the ansatz (\refeq{magnetic-field-ansatze}). 

Let us look at the electric component corresponding to the magnetic field (\refeq{magnetic-field-ansatze}). The Amp\`{e}re law (\refeq{Ampere-gravitation}) determines the evolution of $B^{i} (t,\mathbf{x})$ in terms of metric and of components $E^i (t,\mathbf{x})$ of the electric field. By contemplating the equation (\refeq{Ampere-gravitation}) and the ansatz (\refeq{magnetic-field-ansatze}), we conclude that the electric field should have the following form 
\begin{equation}
E^i (t,\mathbf{x}) = \frac{f(t,\mathbf{x})}{N(t,\mathbf{x})} 
\left[
\left( \mathcal{L}_{\mathbf{m}}  \bar{\phi} (t,\mathbf{x}) \right) D^{i}\phi (t,\mathbf{x}) 
-
\left( \mathcal{L}_{\mathbf{m}}  \phi (t,\mathbf{x}) \right) D^{i} \bar{\phi} (t,\mathbf{x})
\right].
\label{electric-field-ansatze}
\end{equation}
The fields proposed in the equations (\refeq{magnetic-field-ansatze}) and (\refeq{electric-field-ansatze}) must verify the equation of motion 
(\refeq{Ampere-gravitation}). The verification can be done by plugging the fields into the Amp\`{e}re law. Then the most rapid way to prove that the equation (\refeq{Ampere-gravitation}) is satisfied is to show that both left- and right-hand sides of it take the same form, namely,  
\begin{align}
& \varepsilon^{ijk} f 
\left(
\partial_t \partial_j \phi - \beta^r \partial_r \partial_j \phi - \partial_r \phi \partial_j \beta^r
\right)\partial_k \bar{\phi}
\nonumber
\\
& +
\varepsilon^{ijk} f 
\left(
\partial_t \partial_k \bar{\phi} - \beta^r \partial_r \partial_k \bar{\phi} - \partial_r \bar{\phi} \partial_k \beta^r
\right)\partial_j \phi.
\label{common-form-lhs-rhs-Ampere-gravitation}
\end{align}
This result can be proved by direct calculations on each side of the equation. For more details we refer to \cite{Vancea:2017tmx}.

The above analysis lead us to the conclusion that the fields $B^{i}(t,\mathbf{x})$ and $E^{i}(t,\mathbf{x})$ 
given by the relations (\refeq{magnetic-field-ansatze}) and (\refeq{electric-field-ansatze}) are field line solutions of two of the Einstein-Maxwell equations. These solutions satisfy the orthogonality property
\begin{equation}
\gamma^{ij}(t,\mathbf{x})E_i(t,\mathbf{x}) B_j (t,\mathbf{x}) = 0 \, ,
\qquad \forall p \in U_p \, .
\label{E-B-orthogonality}
\end{equation}
As in the flat space-time, the magnetic field lines are the level lines of the complex function $\phi$. However, the interpretation of the electric field in terms of field lines is not transparent in the relation (\refeq{electric-field-ansatze}). That can be remedied by constructing an electric field line solution as was done in the flat space-time. 
An important tool that was used there was the invariance of Maxwell's equations under the duality between the electric and magnetic fields in the vacuum. In the absence of sources, the Einstein-Maxwell equations without sources are invariant under the electric-magnetic duality, too, as it was shown in \cite{Deser:1976iy}. This symmetry allows one to find the line solutions of the Faraday law and the electric Gauss law which are given by the equations (\refeq{Faraday-gravitation}) and (\refeq{Gauss-E-gravitation}), respectively. 

Let us introduce a second complex field $\theta: U_p \rightarrow \mathbb{C}$ and note that the constraint on the electric field is the same as the one on the magnetic field. It follows that one can take the electric field of the same form as the magnetic field given by the equation (\refeq{magnetic-field-ansatze}). Then we can write 
\begin{equation}
E^{i} (t,\mathbf{x}) = g(t,\mathbf{x}) \varepsilon^{ijk} (t,\mathbf{x}) D_{j} \bar{\theta} (t,\mathbf{x}) D_{k} \theta (t,\mathbf{x}),
\label{electric-field-ansatze-1}
\end{equation}
where $g(t,\mathbf{x})$ is an arbitrary real smooth field on $U_p$ that depends on the adapted coordinates only implicitly as $g(\theta (t,\mathbf{x}),\bar{\theta}(t,\mathbf{x}))$. By repeating the arguments given above where we have discussed the electric field line solution, we conclude that the magnetic field line should have the following form
\begin{equation}
B^i (t,\mathbf{x}) = \frac{g(t,\mathbf{x})}{N(t,\mathbf{x})} 
\left[
\left( \mathcal{L}_{\mathbf{m}}  \bar{\theta} (t,\mathbf{x}) \right) D^{i}\theta (t,\mathbf{x}) 
-
\left( \mathcal{L}_{\mathbf{m}}  \theta (t,\mathbf{x}) \right) D^{i} \bar{\theta} (t,\mathbf{x})
\right].
\label{magnetic-field-ansatze-1}
\end{equation}
Since the equation of motion and the constraints for the electric and magnetic fields 
have the same form, the proof that the fields given by the equations (\refeq{electric-field-ansatze-1}) and (\refeq{magnetic-field-ansatze-1}) are solutions of the second set of Einstein-Maxwell equations (\refeq{Gauss-E-gravitation}) and (\refeq{Faraday-gravitation}) is the same as in the case of the magnetic field.

One can easily verify that the electromagnetic duality implies the existence of a relationship between the scalar fields $\phi$ and $\theta$ that must satisfy the following equations simultaneously \cite{Vancea:2017tmx}
\begin{align}
f(\phi,\bar{\phi}) \varepsilon^{ijk} 
D_j \phi  D_k \bar{\phi} 
& = 
\frac{g(\theta,\bar{\theta})}{N} 
\left[
\left( \mathcal{L}_{\mathbf{m}}  \bar{\theta}  \right) D^{i}\theta  
-
\left( \mathcal{L}_{\mathbf{m}}  \theta \right) D^{i} \bar{\theta} 
\right],
\label{f-g-nonlinear-equations-1}
\\
g(\theta,\bar{\theta}) \varepsilon^{ijk}  D_{j} \bar{\theta}  D_{k} \theta 
& =
\frac{f(\phi,\bar{\phi})}{N} 
\left[
\left( \mathcal{L}_{\mathbf{m}}  \bar{\phi} \right) D^{i}\phi  
-
\left( \mathcal{L}_{\mathbf{m}}  \phi \right) D^{i} \bar{\phi} \right].
\label{f-g-nonlinear-equations-2}
\end{align}
These equations form a set of non-linear local constraints on the functions $f$ and $g$ and they must be satisfied at all times. There are no other constraints on $f$ and $g$ which shows that there are actually families of field line solutions rather than isolated solutions. 
One of these solutions is particularly interesting because it is a generalization of the
Ra\~{n}ada field.

\subsection{Local generalization of Ra\~{n}ada's solution}

As discussed in the chapter \cite{Vancea:2019zdl} from this volume, the first field line solution of Maxwell's equations was given by Ra\~{n}ada in \cite{Ranada:1989wc,Ranada:1990}. The Ra\~{n}ada field 
is given in terms of two scalar functions $f$ and $g$ that play the same role as the scalar fields that have shown up in the previous subsection. The main difference between the two cases is that the form of $f$ and $g$ is fixed in the  Ra\~{n}ada field to the following functions
\begin{align}
f &= \frac{1}{2 \pi i} \frac{1}{(1+|\phi|^2)^2},
\label{topological-f}
\\
g &= \frac{1}{2 \pi i} \frac{1}{(1+|\theta|^2)^2}.
\label{topological-g}
\end{align} 
It is interesting to see what is the gravitating electromagnetic field that is obtained by using the functions from the equations (\refeq{topological-f}) and (\refeq{topological-g})
in the vector fields $E^{i}(t,\mathbf{x})$ and $B^{i}(t,\mathbf{x})$ discussed above. By plugging these equations into the electric and magnetic field line solutions obtained in the previous subsection, and after some algebraic manipulation of the vector fields, we obtain the following result
\begin{align}
B^i (t,\mathbf{x}) & = \varepsilon^{ijk} D_j \alpha_1 (t,\mathbf{x}) D_k \alpha_2 (t,\mathbf{x}),
\label{magnetic-field-ansatze-2}
\\
E^i (t,\mathbf{x}) & = \varepsilon^{ijk} D_j \beta_1 (t,\mathbf{x}) D_k \beta_2 (t,\mathbf{x}),
\label{electric-field-ansatze-2}
\end{align}
where $\alpha_1$, $\alpha_2$, $\beta_1$ and $\beta_2$ are real scalar fields related to the complex fields $\phi$ and $\theta$ by the following relations
\begin{align}
\alpha_1 &= \frac{1}{1+|\phi|^2},
\hspace{0.5cm}
\alpha_2 = \frac{\Phi}{2 \pi},
\hspace{0.5cm}
\phi = |\phi| e^{i \Phi},
\label{alpha-phi-identification}
\\
\beta_1 &= \frac{1}{1+|\theta|^2},
\hspace{0.5cm}
\beta_2 = \frac{\Theta}{2 \pi},
\hspace{0.5cm}
\theta = |\theta| e^{i \Theta}.
\label{beta-phi-identification}
\end{align}
The fields given by the equations (\refeq{magnetic-field-ansatze-2}) and (\refeq{electric-field-ansatze-2}) represent the local generalization of the Ra\~{n}ada solution to the hyperbolic space-time $M$. It is important to note that the boundary conditions on $\phi$ and $\theta$ that generate the electromagnetic knots in the Minkowski space-time cannot be imposed automatically on the fields from the equations (\refeq{alpha-phi-identification}) and (\refeq{beta-phi-identification}) since the limit  $\mathbf{\mathbf{x}} \to \infty$ is not well defined in the neighbourhood $U_p$ in general. 

We conclude this section by observing that the gravitating electromagnetic field 
can be analysed in close analogy with the electromagnetic field in flat space-time  
in the $1+3$-formalism. In this way, we can give a simple interpretation to the physical fields in terms of local geometrical and topological quantities. Also, the flat space-time limit can be easily obtained by choosing the Gauss normal coordinates systems with 
$N=1, \pmb{\beta} = 0$. The fields $E^{i}(t,\mathbf{x})$ and $B^{i}(t,\mathbf{x})$ take the form of Ra\~{n}ada's solutions in this frame. This fact leads to the conclusion that the gravitating field lines solutions represent a natural local generalization of Ra\~{n}ada's solutions. As mentioned above, it is possible to obtain global solutions if the topology of $M$  does not impose any obstruction to the existence of global differential 2-forms. 

\section{The Kopi\'{n}ski-Nat\'{a}rio field}

In this section, we will present a particular field line solution that can be extended globally in the \emph{Einstein universe} \cite{Kopinski:2017nvp}. 

Let us recall the line element of the cosmological FRW metric \cite{Weinberg:1972kfs} that is given by the following relation
\begin{equation}
d s^2 = - d t^2 + a(t)^2 
\left[
\frac{d r^2}{ 1 - k r^2} 
+ r^2 \left(
d \theta^2 + \sin^2 (\theta) d \phi^2
\right)
\right] \, .
\label{FRW-metric}
\end{equation}
The parameter $a$ and the variable $r$ can be rescaled such that $k$ take the integer values $-1,0$ or $+1$. If $k=0$, the universe is flat, if $k=+1, \, r = \sin (\chi)$ the universe is closed, and if $k=-1, \, r = \sin \psi $ the universe is open. In what follows, we are interested in the foliated space-time with leaves $\Sigma$ isomorphic to $S^3$. This correspond to a close universe. Matter content can be added to the model such that the homogeneity and the isotropy are preserved. Then the matter content is characterized by the following energy-momentum tensor
\begin{equation}
T_{\mu \nu} = \left( \rho + P \right) u_{\mu} u_{\nu}  + P g_{\mu \nu} 
\, ,
\label{EnMom-tensor-perfect-fluid}
\end{equation}
where $u^{\mu}$ is an unitary time-like vector, $\rho$ is the density of mass and $P$ is the density of pressure. Although $T_{\mu \nu}$ is the energy-momentum tensor of a perfect fluid, many matter models can be put into this form, including the electromagnetic field. In the co-moving frame, one can choose $u^{\mu} = (1,0,0,0)$ and the energy-momentum tensor becomes diagonal ${T^{\mu}}_{\nu}= \mbox{diag}(-\rho, P, P, P)$. Recall the Einstein's equations with a cosmological constant $\Lambda$  
\begin{equation}
G_{\mu \nu} + \Lambda g_{\mu \nu} = 8 \pi T_{\mu \nu} \, .
\label{Einstein-equations}
\end{equation}
If the FRW metric from the equation (\refeq{FRW-metric}) and the energy-momentum tensor from the equation (\refeq{EnMom-tensor-perfect-fluid}) are substituted into the equation (\refeq{Einstein-equations}), the following set of equations is obtained
\begin{align}
3 \frac{\dot{a}^2 + k}{a^2} & = 8 \pi \rho + \Lambda\, ,
\label{Einstein-eq-1}
\\
\frac{2 a \ddot{a} + \dot{a}^2 + k}{a^2}  & = 8 \pi P  + \Lambda\, ,
\label{Einstein-eq-2}
\\
\frac{\ddot{a}}{a^2} & = -\frac{4 \pi}{3} 
\left(
\rho + 3 P
\right)
+ \frac{\Lambda}{3}
 \, .
\label{Einstein-eq-3}
\end{align}
We note that the equations (\refeq{Einstein-eq-1})-(\refeq{Einstein-eq-3}) are not all independent of each other. These are the fundamental equations of cosmology and they describe a large variety of cosmological models that are homogeneous and isotropic. In particular, the \emph{Einstein universe} is characterized by the following constraints
\begin{equation}
\dot{a} = \ddot{a} = 0 \, , \qquad P = 0 \, .
\label{Einstein-universe}
\end{equation}
By using the equation (\refeq{Einstein-universe}) together with the cosmological equations (\refeq{Einstein-eq-1})-(\refeq{Einstein-eq-3}), the following constitutive equations of the Einstein universe are obtained
\begin{equation}
\frac{3k}{a^2} = \Lambda + 8 \pi \rho \, ,
\qquad
\frac{k}{a^2} = \Lambda \, ,
\qquad
k  = 4 \pi a^2 \rho \, .
\label{Einstein-universe-1}
\end{equation}
The Einstein universe is a homogeneous model in space and time, but it is unstable to perturbations $a \to a + \varepsilon$, where $\varepsilon << 1$. Also, for a baryonic matter content, it requires that $k = + 1$, so the universe is closed. One can determine its parameters up to a constant $C$ and the result is given by the following relations
\begin{equation}
a = \frac{3 C}{2} \, ,
\qquad
\Lambda = \frac{4}{9C^2} \, .
\label{Einstein-universe-2}
\end{equation}
The Einstein universe is not physical since the experimental data favours an expanding universe. Nevertheless, it is still an interesting model to be explored from both mathematical and physical point of view due to its highly homogeneous structure. 

It is convenient to represent the Einstein universe in a coordinate system that displays the spherical symmetries of leafs. Since the space-time is foliated and has the topology $M \simeq \mathbb{R} \times S^3$, we can introduce an adapted coordinate system $(t,x^i)$ at any point $p \in M$ and associate to it the canonical basis $(\mathbf{e}_0 = \partial_t, 
\mathbf{e}_i = \partial_i )$ of the tangent space $\mathcal{T}_p (M)$. Then according to  Weyl's postulate, $\mathbf{e}_0$ is orthogonal to the leaf $\Sigma_t \simeq S^3$
\begin{equation}
g(\mathbf{e}_0 \, , \, \mathbf{e}_i ) = 0 \, .
\label{Weyl-postulate-leaf}
\end{equation}
We use the fact that the leaf space can be represent as a group $S^3 \simeq SU(2)$. Then one can choose the basis $(\mathbf{e}_i) \in \mathcal{T}_p (S^3)$ such that the $su(2)$ algebra
is satisfied by its elements
\begin{align}
\mathbf{e}_0 & = \partial_t \in \mathcal{T}_p (\mathbb{R}) \, ,
\label{KN-tetrad-1}
\\
\left[
\mathbf{e}_i \, , \, \mathbf{e}_j
\right]
& = 2 \varepsilon_{ijk} \mathbf{e}_k \, ,
\qquad \mathbf{e}_i
\in \mathcal{T}_p (S^3) \, .
\label{KN-tetrad-2}
\end{align}
Note that the authors of \cite{Kopinski:2017nvp} used the notation $\mathbf{e}_{\mu} = X_{\mu}$ so let us adopt it in what follows. 

The dual basis to $(\mathbf{e}_{\mu})$, denoted by $(\pmb{\theta}^{\mu} )$, has the same decomposition along the directions of the foliated space-time manifold $M$. The elements of the dual basis also obey the $su(2)$ algebra since they satisfy the following equation
\begin{equation}
d \pmb{\theta}^i = - \varepsilon_{ijk} \, d \pmb{\theta}^j \wedge d \pmb{\theta}^k \, . 
\label{KN-tertrad-3}
\end{equation}
One important feature of the Einstein universe is that it is locally conformal equivalent to the Minkowski space-time. By using this equivalence, the local Einstein-Maxwell equations take the same form as the Maxwell equations in some region of the flat space-time. This is possible since the Maxwell equations are invariant under the conformal transformations \cite{Kopinski:2017nvp}. As discussed in the previous sections, they have the following form in the absence of sources
\begin{equation}
d F = 0 \, , \qquad d \star F = 0 \, .
\label{KN-Maxwell-equations}
\end{equation}
The decomposition of the electromagnetic 2-form in to electric and magnetic components is standard and it is given by the following relation
\begin{equation}
F = E^i \, \pmb{\theta}^i \wedge \pmb{\theta}^0 
+ \frac{1}{2} B^i \varepsilon_{ijk} \, \pmb{\theta}^j \wedge \pmb{\theta}^k \, .
\label{KN-F-decomposition}
\end{equation}
By plugging the equation (\refeq{KN-F-decomposition}) into the equations (\refeq{KN-Maxwell-equations}), one can easily obtain the following set of equations
\begin{align}
& X_i (E^i )  = X_i (B^i ) = 0 \, ,
\label{KN-eq-1}
\\
& \dot{B}^i  - 2 E^i + \varepsilon_{ijk} X_j (E^k)  = 0 \, ,
\label{KN-eq-2}
\\
& \dot{E}^i  + 2 B^i - \varepsilon_{ijk} X_j (B^k)  = 0 \, .
\label{KN-eq-3}
\end{align}
The Kopi\'{n}ski-Nat\'{a}rio field is obtain by making the ansatz that the components of the electric and magnetic fields depend only on the time as measured in a stationary frame \cite{Kopinski:2017nvp}. This assumption is valid in the Einstein universe which is a stationary closed model. As such, the variations along the directions of the $SU(2)$ manifold vanish and the equations (\refeq{KN-eq-2}) and (\refeq{KN-eq-3}) take the following form
\begin{align}
& \dot{B}^i  - 2 E^i = 0 \, ,
\label{KN-eq-2-a}
\\
& \dot{E}^i  + 2 B^i  = 0 \, .
\label{KN-eq-3-a}
\end{align}
The simplest solution of the equations (\refeq{KN-eq-2-a}) and (\refeq{KN-eq-3-a}) obtained in \cite{Kopinski:2017nvp} has the following form
\begin{align}
E (t) & = E_0 \cos (2t) X_1 \, ,
\label{KN-Sol-E}
\\
B (t) & = B_0 \cos (2t) X_1 \, .
\label{KN-Sol-B}
\end{align}
However, one can see that the equations (\refeq{KN-eq-2-a}) and (\refeq{KN-eq-3-a}) 
admit many other solutions generated by the second degree differential equation
\begin{equation}
\ddot{f} \pm 4 f  = 0 \, .
\label{KN-2-deg-diff-eq}
\end{equation}
One important remark is that the components of the Kopi\'{n}ski-Nat\'{a}rio field are not orthogonal to each other as they share the same direction of $SU(3)$. Also, their field lines are unstable under small perturbations. 

As the authors noted in their paper \cite{Kopinski:2017nvp}, the Einstein universe and the Minkowski space-time are conformally equivalent to each other in some open regions, which allows one to conformally map the objects defined in the Einstein universe into similar objects that live in the Minkowski space-time. The conformal mapping is defined by the following transformation of the metric tensor
\begin{equation}
g_{\mu \nu} = \Omega \, \eta_{\mu \nu} \, ,
\label{KN-conformal-mapping}
\end{equation}
where $\eta_{\mu \nu}$ is the metric of Minkowski space-time and $\Omega$ is the conformal factor. In particular, if one interprets the energy of the field line solution in terms of the energy defined in the flat space-time, the following relation is found \cite{Kopinski:2017nvp}
\begin{equation}
\mathcal{E} (t) = \frac{E^{2}_{0}}{12}
\left[
9 \pi \sin (t) + \pi \sin (3t) + 12 \pi^2 \cos (t) - 12 \pi t \cos (t)
\right] \, ,
\label{KN-energy-flat}
\end{equation}
The energy $\mathcal{E} (t)$ shows that the Kopi\'{n}ski-Nat\'{a}rio solution describes a radiation field. 

In the reference \cite{Kopinski:2017nvp}, similar considerations were made for a general FRW closed universe described by the line element from the equation (\refeq{FRW-metric}). In that case, the vector fields corresponding to the metric are given by the following relations
\begin{equation}
\bar{\mathbf{X}}_0  = \partial_t \, ,
\qquad \bar{\mathbf{X}}_i  = a^{-1}(t) \partial_i
\, .
\label{KN-vectors-FRW}
\end{equation}
The dual basis associated to them is composed by the following tetrad fields
\begin{equation}
\bar{\mathbf{\theta}}^0  = dt \, ,
\qquad \bar{\mathbf{\theta}}^i  = a(t) \bar{\mathbf{\theta}}^i
\, ,
\label{KN-tetrads-FRW}
\end{equation} 
The electromagnetic 2-form field can be decomposed with respect to this basis and the following relation is obtained
\begin{equation}
F = - E^1 \, \bar{\mathbf{\theta}}^0 \wedge \bar{\mathbf{\theta}}^1
+ B^1\, \bar{\mathbf{\theta}}^2 \wedge \bar{\mathbf{\theta}}^3 \, .
\label{KN-F-FRW}
\end{equation} 
Then the Einstein-Maxwell equations in the vacuum take the following form
\begin{align}
\frac{d}{dt} \left[ a^2(t) B^1 \right] & = 2a E^1 \, ,
\label{KN-EM-FRW-1} 
\\
\frac{d}{dt} \left[ a^2(t) E^1 \right] & = - 2a B^1 \, ,
\label{KN-EM-FRW-2}
\end{align} 
The solution of the equations (\refeq{KN-EM-FRW-1}) and (\refeq{KN-EM-FRW-2}) can be easily obtained by integration. The result is given by the following relations
\begin{align}
E^1 (t) & = \frac{E_0}{a^2 (t)} \cos \left[ \int dt \, a^{-1} (t) \right] \, ,
\label{KN-EM-FRW-1a} 
\\
B^1 (t) & = \frac{E_0}{a^2 (t)} \sin \left[ \int dt \, a^{-1} (t) \right] \, .
\label{KN-EM-FRW-2a}
\end{align}
The energy of this solution goes as $\sim E^{2}_{0}a^{-4}(t)$. 
The Einstein universe represents a particular case of the above equations for which $a$ is constant. For a more complete discussion of the properties of this solution we refer the reader to the original work \cite{Kopinski:2017nvp}.

\section*{Appendix}

In this appendix, we collect some definitions and properties of the Lie derivative, the interior derivative and the extrinsic curvature that have been used in the text. This is a  standard material and can be found in any classic text on differential geometry. For applications in physics, see e. g. \cite{Frankel:1997ec}.

In all the definitions reviewed here, we will consider a differentiable manifold $M$ of dimension $\mbox{dim} (M) = n$.
\begin{defn}
If $T \in \mathcal{T}^{p}_{q}(M)$ is a tensor field of rank $(p,q)$ and $X \in \mathcal{X} (M)$ is a differentiable vector field, then the \emph{Lie derivative} of $T$ along $X$ is defined as follows
\begin{equation}
({\mathcal{L}}_{Y} T)_{p} = 
\left. 
\frac {d}{dt} 
\right|_{t=0}
\left[
(\mu_{-t})_{*} T_{\varphi _{t}(p)}
\right]
=
\left.
\frac {d}{dt}
\right|_{t=0}
\left[
(\mu_{t})^{*}T_{p}
\right].
\label{Lie-derivative-definition}
\end{equation}
Here, $\mu : I \times M \to M$, $I \subset \mathbb{R}$ is the one-parameter semigroup of diffeomorphisms on $M$ generated by the flow of $X$ with the action
\begin{equation}
x \to \mu_t (x) = \mu(t,x) \, , 
\, \,
\forall x \in M \, ,
\qquad
X (x) = \left. 
\frac {d \mu (t,x)}{dt} 
\right|_{t=0} \, .
\label{One-parameter-flow}
\end{equation} 
\end{defn}
The Lie derivative obeys the following axioms 
\begin{align}
{\mathcal{L}}_{X} f& = X(f) \, ,
\label{Lie-derivative-axiom-1}
\\
[{\mathcal{L}}_{X} \, , \, d \,] & = 0 \, , 
\label{Lie-derivative-axiom-2}
\\
{\mathcal{L}}_{X}(T \otimes S) & =
({\mathcal{L}}_{X}T ) \otimes S + T \otimes ({\mathcal{L}}_{X}S)
\, , 
\label{Lie-derivative-axiom-3}
\\
{\mathcal{L}}_{X}
\left(T(X_{1},\ldots ,X_{n}
\right) & =
\left({\mathcal {L}}_{X}T
\right)
\left( X_{1},\ldots ,X_{n}
\right) 
\nonumber
\\
& + 
T\left(
{\mathcal {L}}_{X}X_{1}
,\ldots ,X_{n}
\right)
+\cdots +
T\left(X_{1},\ldots ,{\mathcal {L}}_{X} X_{n}
\right) \, ,
\label{Lie-derivative-axiom-4}
\end{align}
where $T$ and $S$ are arbitrary tensor fields and $d$ is the exterior derivative. In the above relations, the objects considered there have natural properties. It is easy to show that the following relations hold
\begin{align}
{\mathcal{L}}_{X} Y(f) &= X(Y(f))-Y(X(f)) = [X , Y] (f) \, ,
\label{Lie-derivative-properties-1}
\\
{\mathcal{L}}_{X} \omega & = \iota_{X} d \omega +d \iota_{X}\omega \, .
\label{Lie-derivative-properties-2}
\end{align}
In the equation (\refeq{Lie-derivative-properties-2}), we have denoted by $\omega$ a differential form on $M$ and by $\iota$ the interior product to be defined below.
\begin{defn}
Let $X \in \mathcal{X} (M)$ be a smooth vector field on $M$ and $\omega \in \Omega^{k} (M)$ a $k$-form. Then the \emph{interior product} of $X$ and $\omega$ is a $(k-1)$-form $\iota_X \omega$ defined by the following property
\begin{equation}
\iota_X \omega (X_1 , \ldots , X_{k-1} ) = \omega (X , X_1 , \ldots X_{k-1} ) \, .
\label{interior-product-def}
\end{equation}
If $k=0$ we take by definition $\iota_{X} \omega = 0$.
\end{defn}
It is easy to show that the following equalities are true
\begin{align}
\iota_X(\omega \wedge \gamma) &= \iota_X \omega  \wedge \gamma
+(-1)^k \omega \wedge \iota_X \gamma \, ,
\, \,
\forall \omega \in \Omega^{k} (M) \, ,
\forall \gamma \in \Omega^{q} (M) \, ,
\label{interior-product-properties-1}
\\
\iota_{[X,Y]} &= 
\left[
\mathcal{L}_{X} , \iota_{Y}
\right] \, ,
\qquad
\forall X , Y \in \mathcal{X} (M) \, ,
\label{interior-product-properties-2}
\\
\iota_{X} \iota_{Y} \omega & = -\iota_{Y} \iota_{X} \omega \, ,
\, \, 
\forall X , Y \in \mathcal{X} (M) \, , \forall \omega \in \Omega^{k} (M)
\, .
\label{interior-product-properties-3}
\end{align}
Let us we review the \emph{extrinsic curvature} of an embedded surface in three dimensions. The generalization to four-dimensions is straightforward. Consider an oriented surface $\Sigma \in \mathbb{R}^{3}$ on which the orientation is defined by the unit vector field $\mathbf{n}$. Then one can define the \emph{Gauss map} as follows
\begin{equation}
\pmb{\nu} \, : \, \Sigma \to S^2 \, ,
\qquad
\pmb{\nu}(x) = \mathbf{n} (x) \, .
\label{Weingarten-map}
\end{equation}
Since $\pmb{\nu}$ is smooth, it induces the following map
\begin{equation}
D_{x}\pmb{\nu} \, : \, \mathcal{T}_{x} (\Sigma ) \to 
\mathcal{T}_{\pmb{\nu} (x)} (S^2 ) \, .
\label{derivative-map}
\end{equation}
We know that $\mathcal{T}_{\pmb{\nu} (x)} (S^2 ) \simeq \mathcal{T}_{x} (\Sigma ) $. Then the derivative $D_{x}\pmb{\nu}$ defines the \emph{Weingarten map} as follows
\begin{equation}
\mathcal{W}_{x} = - D_{x}\pmb{\nu} \, : \, \mathcal{T}_{x} (\Sigma ) \to 
\mathcal{T}_{x} (\Sigma ) \, .
\label{Weingarten-map-def}
\end{equation}
The symmetric bilinear two-form curvature map is obtained from $\mathcal{W}_{x}$ as follows 
\begin{equation}
K_{x} (X,Y) = \langle \mathcal{W}_{x} X, Y \rangle \,
\label{curvature-Weigarten-bilinear-form}
\end{equation}
for any $X, Y \in \mathcal{T}_{x} (\Sigma )$.


\newpage

\begin{thebibliography}{Proper}


\bibitem{Trautman:1977im} 
  Trautman, A. (1977). ``Solutions of the Maxwell and Yang-Mills Equations Associated with Hopf Fibrings,''
  \emph{Int. J. Theor. Phys.} 16, 561.

\bibitem{Ranada:1989wc} 
  Ra\~{n}ada, A. F. (1989).
  ``A Topological Theory of the Electromagnetic Field,''
  \emph{Lett. Math. Phys.}  18, 97.

\bibitem{Ranada:1990}
  Ra\~{n}ada,  A. F. (1990).
``Knotted solutions of the Maxwell equations in vacuum,''
  \emph{J. Phys. A} 23, L815. 


\bibitem{Arrayas:2017sfq} 
  Array\'{a}s, M., Bouwmeester, D. and Trueba, J. L. (2017).
  ``Knots in electromagnetism,''
  \emph{Phys. Rept.} 667.

\bibitem{Vancea:2019zdl} 
  I.~V.~Vancea,
  ``Knots and the Maxwell's Equations,''
  in \emph{An Essential Guide to Maxwell's Equations}, Ed. Casey
   Erickson, Nova Science Publishers.



\bibitem{Dalhuisen:2012zz} 
  Dalhuisen, J. W. and Bouwmeester, D. (2012).
  ``Twistors and electromagnetic knots,''
  \emph{J. Phys. A} 45, 135201.

\bibitem{Swearngin:2013sks} 
Swearngin, J., Thompson, A., Wickes, A., Dalhuisen, J. W. and Bouwmeester, D. (2013). 
  ``Gravitational Hopfions,''
  arXiv:1302.1431 [gr-qc].

\bibitem{Thompson:2014pta} 
Thompson, A., Swearngin, J. and Bouwmeester, D. (2014).
 `Linked and Knotted Gravitational Radiation,''
  \emph{J. Phys. A} 47, 355205.

\bibitem{Thompson:2014owa}
Thompson, A., Wickes, A., Swearngin, J. and Bouwmeester, D. (2015). 
  ``Classification of Electromagnetic and Gravitational Hopfions by Algebraic Type,''
  \emph{J. Phys. A} 48, 205202.

\bibitem{Alves:2018wku} 
Alves, D. W. F. and Nastase, H. (2018).
 ``Hopfion solutions in gravity and a null fluid/gravity conjecture,''
  arXiv:1812.08630 [hep-th].


\bibitem{Vancea:2017tmx} 
  Vancea, I. V. (2017).
  ``On the existence of the field line solutions of the Einstein -
  Maxwell equations,''
  \emph{Int. J. Geom. Meth. Mod. Phys.} 15, 1850054.

\bibitem{Kopinski:2017nvp} 
  Kopi\'{n}ski, J. and Nat\'{a}rio, J. (2017).
  ``On a remarkable electromagnetic field in the Einstein Universe,''
  \emph{Gen. Rel. Grav.} 49, 81.

\bibitem{Silva:2018ule} 
  Costa e Silva, W., Goulart, E. and Ottoni, J. E. (2018).
  ``On spacetime foliations and electromagnetic knots,''
  arXiv:1809.09259 [math-ph].


\bibitem{Lechtenfeld:2017tif} 
  O.~Lechtenfeld and G.~Zhilin,
  ``A new construction of rational electromagnetic knots,''
  Phys.\ Lett.\ A \textbf{382}, 1528 (2018).
  \doi{10.1016/j.physleta.2018.04.027}

\bibitem{Kumar:2020xjr} 
  O.~Lechtenfeld and K.~Kumar,
  ``On rational electromagnetic fields,''
  arXiv:2002.01005 [hep-th].



\bibitem{Gourgoulhon:2012} 
  Gourgoulhon, \'{E}. (2012).
  ``3+1 Formalism in General Relativity: Bases of Numerical Relativity,''
  Lecture Notes in Physics, Volume 846,
  Springer-Verlag.
  
\bibitem{Weinberg:1972kfs} 
  Weinberg, S. (1972).
  ``Gravitation and Cosmology : Principles and Applications of the General Theory of Relativity,'' John Wiley and Sons.


\bibitem{Jackson:1998nia} 
  Jackson, J. D. (1962).
  ``Classical Electrodynamics,''
  John Wiley and Sons.

\bibitem{Frankel:1997ec} 
  Frankel, T. (1997).
  ``The geometry of physics: An introduction,''
  Cambridge University Press.


\bibitem{Deser:1976iy} 
  Deser, S. and Teitelboim, C. (1976).
  ``Duality Transformations of Abelian and Nonabelian Gauge Fields,''
  \emph{Phys. Rev. D} 13, 1592.

\end{thebibliography}
\end{document}